\documentclass[graybox, nosecnum]{svmult}

\usepackage{mathptmx}       
\usepackage{helvet}         
\usepackage{courier}        
\usepackage{type1cm}        
\usepackage{makeidx}         
\usepackage{graphicx}        
\usepackage{multicol}        
\usepackage[bottom]{footmisc}
\usepackage{hyperref}        
\usepackage{soul}            

\usepackage{tabularx,pbox} 

\hypersetup{colorlinks=true,urlcolor=blue}
\usepackage[square,numbers]{natbib}

\def\nat{\ref@jnl{Nature}}              
\def\aaps{\ref@jnl{A\&AS}}              
\def\apj{\ref@jnl{ApJ}}                 
\def\aap{\ref@jnl{A\&A}}                
\def\jcap{\ref@jnl{JCAP}}                
\def\mnras{\ref@jnl{MNRAS}}                
\def\prd{\ref@jnl{PRD}}                
\def\aapr{\ref@jnl{AAPR}}                
\def\pasp{\ref@jnl{PASP}}                
\def\physrep{\ref@jnl{PhysRep}}                
\def\prl{\ref@jnl{PhysRevLett}}
\def\ssr{\ref@jnl{SSR}}
\def\caa{\ref@jnl{ChA\&A}}                

\makeindex             

\begin{document}
\title*{X- and Gamma-ray astrophysics in the era of Multi-messenger astronomy}
\author{G. Stratta \thanks{corresponding author} and A. Santangelo}
\institute{G. Stratta \at INAF-IAPS, Via Fosso Del Cavaliere 100, Rome, Italy - INFN-Roma, P.zzale A. Moro, Rome, Italy - INAF/OAS, Via Gobetti 93/D, Bologna, Italy \email{giulia.stratta@inaf.it}
\and A. Santangelo \at Institute of Astronomy and Astrophysics, University of T\"ubingen, Sand 1, 72076, T\"uebingen, Germany \email{andrea.santangelo@uni-tuebingen.de}}
%
%
\maketitle
%
\abstract{
Multi-messenger astronomy is becoming a major avenue to explore the Universe. Several well known astrophysical sources are also expected to emit other 'messenger' than photons: namely cosmic rays, gravitational waves and neutrinos. These additional messengers bring complementary pieces of information to the ones carried by electromagnetic radiation and concur to draw a complete phenomenological picture of several astrophysical events as well as to measure key cosmological parameters. 
Indeed, it is widely believed in the astronomical community that several aspects of fundamental physics and cosmology will be unveiled only within the framework of multi-messenger astronomy. The most recent breakthrough discoveries of a gravitational wave source  associated with a short gamma-ray burst, and of a neutrino event found to be spatially consistent with a flaring blazar, have already shown the key role that high-energy sources will play in multi-messenger observations. The first part of this chapter provides a description of the main properties of gamma-ray bursts, blazars, and other high-energy sources from which we expect to detect gravitational waves and/or neutrinos in the next years, and the achievements that will be reached from  multi-messenger observations.  
The second part of the chapter is focused on the major facilities that are playing and that will play a crucial role in multi-messenger observational campaigns. More specifically, we provide an overview of current and next generation ground-based gravitational wave interferometers and neutrino telescopes, as well as the major X-ray and gamma-ray observatories that will be crucial for multi-messenger observations in the coming years.}

\section{Keywords} 
gamma-ray bursts; blazars; gravitational waves; neutrinos

\section{Introduction}

For millennia, mankind observed the sky by collecting electromagnetic radiation from astrophysical sources, first with the naked eye and then with more and more sophisticated instruments with increased sensitivity and spectral coverage, ultimately disclosing the Universe from radio to gamma-rays up to its infancy. Revolutionary advances in theoretical physics and observational astronomy over the past century provided the basis for modern multi-messenger astronomy. 

In 1912, radiation once thought to be of terrestrial origin was detected on a balloon flight 5300 meters above ground by Victor Hess. The discovery of the cosmic origin of this radiation allowed for its subsequent name: Cosmic Rays (CRs). CRs are high-energy particles arriving from outer space, mainly consisting of protons ($\sim89$\%), i.e. nuclei of hydrogen, the lightest and most common element in the Universe, but also of helium nuclei ($\sim10$\%) and a tiny fraction of heavier nuclei. Their energies span several orders of magnitudes from $\sim10^9$ eV up to $\sim10^{21} $eV. At their arrival at Earth, CRs collide with the nuclei of atoms in the upper atmosphere and generate an electromagnetic cascade. Before entering the atmosphere, cosmic rays are deviated by cosmic magnetic fields during their travel towards the Earth, preventing us from locating the astrophysical source. In any case, their existence and extreme energy point to the presence of cosmic sources, acting as powerful particle accelerators, the identification of which in one of the hot topics of modern high-energy astrophysics. 

During the same epoch, Albert Einstein reached his final formulation of general relativity, and in 1916 he predicted the existence of gravitational waves (GWs) \citep{Einstein1916,Einstein1918}. GWs are space-time perturbations produced by accelerated mass quadrupoles, i.e. accelerated mass with a non axisymmetric distribution, with strength proportional to the second derivative of quadrupole and inversely proportional to the source distance. The most straightfoward astrophysical example of a GW source is a binary star system that, due to the energy dissipation caused by GW emission, gradually decreases the star orbital separation and, consequently, increases the angular velocity. The predicted change of the orbital parameters was observed for the first time in the evolution of the pulsar binary system PSR 1913+16 discovered by Hulse and Taylor in 1974 \cite{Hulse1974,Hulse1975}. However, the very first direct detection of GWs was achieved only on 2015, i.e. one century from their prediction. Contrary to CRs, GWs propagate straight to Earth without changing their travel direction, and interact little with matter during their travel. Their properties encode precious information of the emitting source (e.g. mass, spin, geometry, etc.) complementary to those obtained from observations in the electromagnetic (EM) spectrum. On the other side, GW interferometers have poor sky localization capabilities, and  only through EM observations it is possible to identify the GW source through accurate sky position refinement. 

The last century also witnessed decisive discoveries in neutrino astronomy. Following the first detection of neutrinos from our Sun in 1968, in 1987 about 20 neutrinos were detected at MeV from the core-collapse supernova 1987A in the Large Magellanic Cloud. Recent years marked the first breakthrough detection of cosmic neutrinos from beyond the Local Group environment. Neutrinos are the most abundant elementary particles in the Universe after photons and played a fundamental role in the large structure formation. Their chargeless, stable, and weakly interacting nature allows them to cross large fractions of the Universe as well as to emerge from the innermost and densest regions almost undisturbed, similarly to GWs and unlike CRs. They are thought to be produced in large quantities inside the core of highly energetic astrophysical sources as during the gravitational collapse of massive stars or in the densest regions of active galactic nuclei, where particle acceleration mechanisms are at play and from which high-energy photons are typically absorbed. Therefore, neutrinos serve as better messengers than photons to gain a direct insight of the most distant and/or dense cosmic accelerators.


During the last decade, the advent of sensitive gravitational wave and neutrino detectors, coupled with wide-field, high cadence electromagnetic time-domain surveys, is providing the basis for the 'golden era' of multi-messenger astrophysics through joint GW/neutrino/photon observations from the same source. 

\begin{figure}
    \centering
    \includegraphics[scale=0.5]{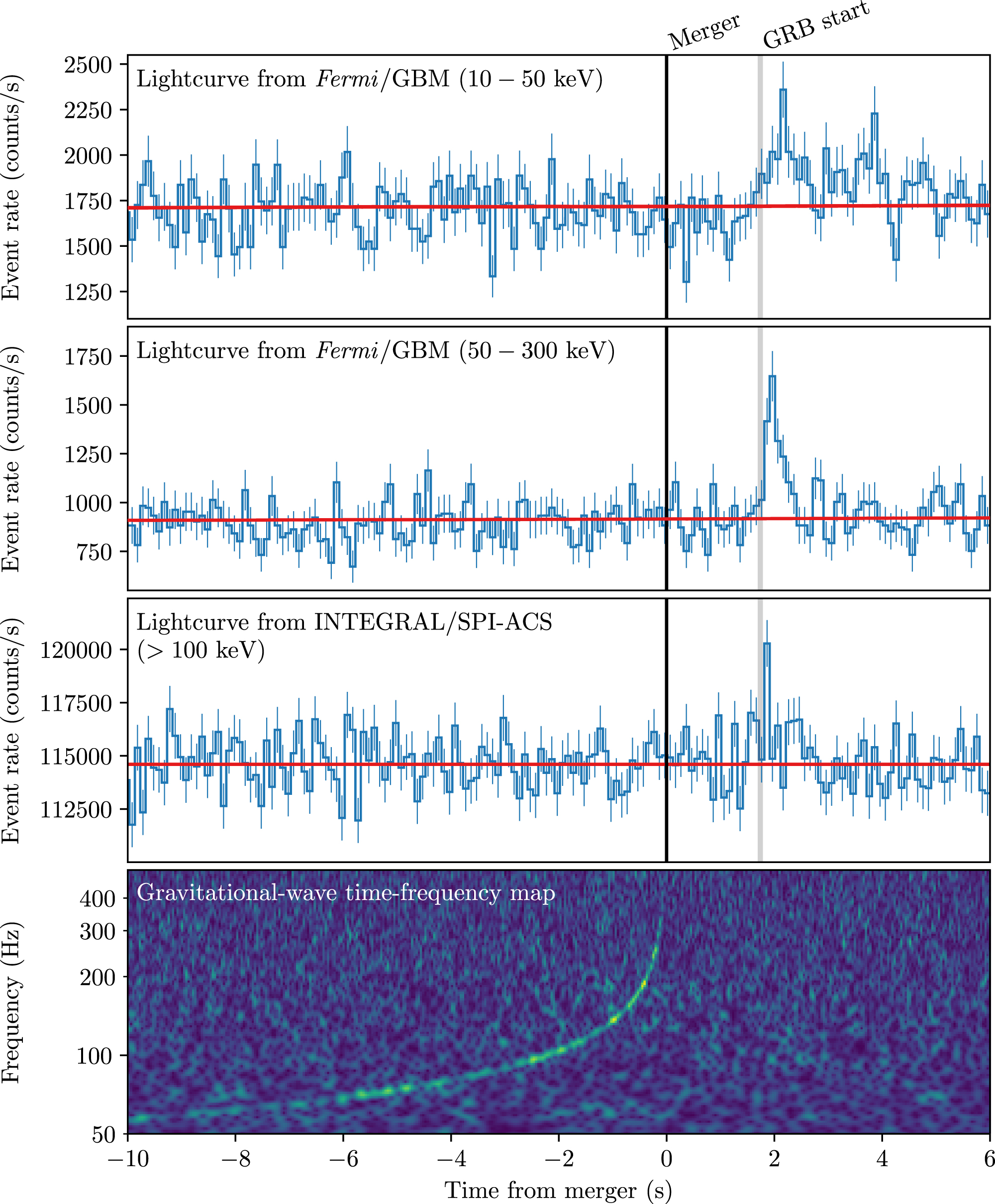}
    \caption{Joint multi-messenger detection of GW170817 and GRB 170817A. The Fermi/GBM lightcurves of GRB 170817A (10-50 keV and 50–300 keV energy range) are plotted on the top two panels, followed by the INTEGRAL/SPI-ACS lightcurve (100 keV-80 MeV). At the bottom the time-frequency map of GW170817 obtained by coherently combining LIGO-Hanford and LIGO- Livingston data si plotted. Figure from Ref. \citep{Abbott2017grbgw} under the terms of the Creative Commons Attribution 3.0 License.}
    \label{fig:gwgrb}
\end{figure}

In 2015, the Advanced Laser Interferometer GW Observatory in the US (aLIGO, \cite{aLIGO}) observed, for the first time, the space-time perturbation generated by the coalescence of a binary system formed by two stellar-mass black holes at a distance of 410 Mpc away from the Earth \cite{LVC-BBH1}. In 2017, the GW interferometer network formed by aLIGO and Advanced Virgo (AdV, \citep{AdVirgo}, detected the first multi-messenger GW source, the merger of a binary neutron star system (GW170817) \cite{Abbott2017gw170817}. Indeed, at the merger epoch and in a sky position consistent with the one obtained from GW data analysis, a short gamma-ray burst (GRB 170817A) was detected (Fig. \ref{fig:gwgrb}), subsequently triggering a massive follow-up observational campaign at all wavelengths. Accurate GW and EM data analysis ultimately permitted to identify as the EM counterpart of GW170817 \cite{Abbott2017grbgw}.

By the same epoch, observations performed with the IceCube neutrino detector \citep{IceCube} allowed us, for the first time, to identify a diffuse flux of very-high-energy neutrinos (10 TeV-10 PeV) consistent with being of astrophysical origin, though the neutrino sources are still unidentified \cite{Aarsten2013}. In one case, a neutrino source could be recognized as a flaring blazar TXS0506+056 at redshift z=0.34 \cite{Aartsen2018}. 

Multi-messenger transients (e.g. GRB, cc-SNe, flaring blazars, etc.) detected in the EM spectrum, can be used for the increase in the sensitivity of low signal-to-noise GW and neutrino events by setting the event epoch and sky position as priors, thus reducing the parameter space in the data analysis. On the other hand, once a GW or neutrino event is detected, follow-up observations with ground- and space-based telescopes can be used to detect the EM counterpart,  ultimately providing the source identification and its full characterization.  
{\it In this regard, the high-energy and transient nature of the two cosmic multi-messenger sources detected so far (i.e. the short GRB 170817A and the flaring blazar TXS0506+056), clearly highlight the crucial role of future X-ray and gamma-ray surveys in multi-messenger astrophysics.}

This chapter is devoted to the high-energy observational astrophysics in the era of multi-messenger astronomy and it is structured in two parts. In the first part, we provide a description of the most promising X-ray and gamma-ray sources that we expect to detect jointly with GWs and/or $\nu$s in the next years, together with their main source properties in each emitting domain and major achievements that can be reached through multi-messenger observations. In the second part, we give a description of the main facilities that will allow us to perform observations of GWs, neutrinos together with the most promising high-energy survey facilities as well as X-ray narrow-field telescopes.  

\section{\textit{X-ray and gamma-ray multi-messenger sources}}

The gravitational wave detectors currently operative, and that will provide GW signals up to the 2030s, are ground based km-scale laser interferometers that are sensitive to GWs in the range of 1 to a few thousand Hz \citep{GWIC2020}. At these frequencies, we expect to detect compact binary coalescences (CBCs), i.e. binary systems formed by compact objects as stellar mass black holes (BH) or neutron stars (NS), as well as core-collapse stars and possibly isolated NSs, although the expected GW amplitude is orders of magnitude less with respect to the CBCs. When a CBC has at least one NS, we expect the emission of EM radiation. This can be very bright for the case of short GRBs. On the other hand, core collapse SNe and NSs are known to produce bright X-ray and gamma-ray emission in the form of long GRBs, SN shock break-out, and Soft Gamma Repeaters (SGRs).

High-energy neutrino detectors, such as IceCube, and their future upgrades are approaching the requested sensitivity to routinely detect cosmic neutrinos, and so far the only identified source of a neutrino event belongs to the blazar class of sources. 

On the basis of these considerations, and given the fact that so far the only two GW and neutrino sources detected in the EM spectrum are a gamma-ray burst and a blazar, respectively, it is natural to rank these two classes of high-energy sources among the most promising ones for current and future multi-messenger observational campaigns. 

\subsection{Gamma-ray bursts}


GRBs history starts during the Cold War, when the military satellites VELA serendipitously detected intense flashes of gamma-rays \citep{Klebesadel1973} coming from the sky and identified to be of astrophysical nature. After a long period of sparse observations with small gamma-ray detectors equipped on board of several space missions launched during the '80s, the first leap forward in our understanding of the GRB nature was achieved with the large statistics of events accumulated during the 1990s with the observations by the Burst And Transient Experiment (BATSE) detector \citep{BATSE} on board the Compton Gamma-ray Observatory. BATSE was an all sky monitor sensitive in the band $\sim20-600$ keV with location accuracy of the order of few degrees at best. 


The first breakthrough finding was the isotropic distribution of GRB projected sky coordinates \citep{Briggs1996} that strongly supported the cosmological origin of GRBs and ruled out any Galactic source hypothesis. At cosmological distances, the luminosity inferred by the observed fluxes is of the order of $\sim10^{50}-10^{52}$ erg s$^{-1}$, ranking these events among the most powerful in the whole Universe. 

The second advancement obtained with BATSE observations concerned the GRB burst duration that clearly showed a bimodal distribution, with two peaks at $\sim0.3$ s and $\sim30$ s \citep{Kouveliotou1993}. This result suggested the existence of two different classes of GRBs, identified as 'short' and 'long' GRBs. Later on, this heuristic classification was proved to reflect the real existence of two distinct classes of GRB progenitors. 

By the end of the '90s, the satellite for X-ray Astronomy BeppoSAX \citep{BeppoSAX} allowed for the first time the detection the GRB 'afterglow' in X-rays, a bright non-thermal emission rapidly fading within the first minutes-hours after the burst trigger epoch \citep{Costa1997}. In about $\sim$ 50\% GRBs, the optical afterglow counterpart is also detected, with a longer life time with respect to the X-rays (up to days) and a small fraction is also detected in the radio bands on even longer epochs ($\sim$ several days).

The afterglow discovery represented a crucial advance in the comprehension of the physics at play in GRBs. Indeed, the afterglow emission was theoretically predicted by the so called ’’fireball'' model \citep{Goodman1986,Paczynski1986}.  
The basic concepts of the afterglow originates from what has been labelled as the 'compactness' problem. Indeed, the large opacity $\tau_{\gamma\gamma}$ due to photon-photon interactions, obtained from the observed gamma-ray burst fluences ($\sim10^{-8}-10^{-7}$ erg cm$^{-2}$) and the small source dimension scales estimated from the typical ms-scale light curve variability, should produce thermal GRB spectra, in contrast with the observations. The fireball model introduces the hypothesis that the observed gamma-ray radiation originates from the reprocessing of the kinetic energy of a relativistic expanding ejecta \citep{Shemi1990,Paczynski1990}. The relativistic motion of the source towards the observer allows us to decrease the true gamma-ray photon density, blue-shift the emitted photons by a factor $\Gamma$, and at the same time, to relax the constraints on the source dimensions by $\Gamma^2$, ultimately lowering $\tau_{\gamma\gamma}$, where $\Gamma$ is the bulk Lorentz factor of the expanding source. Typical GRB observed properties matched well with this model by assuming $\Gamma > 100$. Within this paradigm, the  afterglow emission originates from the impact of the relativistic ejecta with the surrounding interstellar matter. The observed non-thermal emission is explained as synchrotron radiation from shock-accelerated electrons interacting with local magnetic fields (e.g. \citep{Sari1998}). 


From an observational point of view, the much longer duration of the afterglow ($\sim$ days) with respect to the gamma-ray bursts ($\sim$ 0.1-100 s) and its slow evolution towards longer wavelengths, enable to perform follow-up observational campaigns with optical and radio telescopes and to refine the sky localization of GRBs\footnote{Gamma-ray detectors have sky localization accuracy that goes from few arcmin up to several degrees} to the arcsecond level or better, that is the precision level required for the identification of the host galaxy. The host galaxy properties encode crucial information on the nature of the GRB progenitors. For instance, long GRBs (LGRBs) hosts are late types and have on average high specific star forming rates as expected if progenitors were young massive stars. On the other hand, short GRB host galaxies are of mixed types, with almost $\sim$50$\%$ of early type, and the GRB position is often coincident with the outermost galaxy regions, suggesting an association with an old stellar population that had enough time to move away from its forming region 
\citep{Berger2014}. Spectroscopic observations of the hosts and/or of the optical afterglows enable us to identify line systems from which it is possible to measure the cosmological redshift of the GRBs. Once the distance is known, fundamental properties, such as the total energy emitted in gamma-rays can be computed.

 Independent of their progenitor nature, both long and short GRBs are interpreted by invoking the same mechanism, i.e. the formation of an accreting compact object as a BH or a NS, forming a relativistic jet from which the burst originates. The burst duration of long GRBs is consistent with an accreting mass reservoir compatible with massive stars (e.g. \citep{Woosley2012}). This was definitively confirmed with the detection of core-collapse supernovae of type Ib,c spatially and temporally consistent with LGRBs (e.g. \citep{Galama1998}). On the other hand, the most accredited interpretation for short GRB progenitors is the coalescence of compact objects as two neutron stars (NS-NS) or a stellar-mass black hole and a neutron star (NS-BH). Direct evidence of this scenario was obtained for the first time with the simultaneous detection of gravitational waves from the NS-NS merger GW170817 and the short GRB 170817A, an event that happened during the summer of 2017, which we describe in the next section.

\subsubsection{Joint GW and EM observations of GRBs}


So far, the only short GRB for which GWs were detected is GRB 170817A. The GW signal was observed with the aLIGO and Advanced Virgo (AdV) network \citep{Abbott2017gw170817}, while the GRB was independently detected with the anticoincidence shields (ACS) of the gamma-ray spectrometer (SPI) on board the INTEGRAL observatory \citep{INTEGRAL}, and with the Gamma-Ray Burst Monitor on board the Fermi high-energy mission \citep{Meegan2009-FermiGBM}. 
The GRB peak flux epoch was 1.7 s after the NS-NS merger epoch \citep{Abbott2017grbgw} (Fig.1). 


The gravitational wave signal of GW170817 was found to be consistent with the predictions from General Relativity for a merging NS-NS system \citep{Abbott2017gw170817}.  
The signal was detected for about 100 s. At early times, the chirp-like time evolution of the frequency is determined primarily by the chirp mass $M_c= \frac{(m1m2)^{3/5}}{(m1+m2)^{1/5}}$, where $m_1$ and $m_2$ are the binary component masses. The chirp mass $M_c$, together with the inclination angle $i$ between the normal to the binary orbital plane and our line of sight, as well as the measured GW amplitude, allow us to directly compute the luminosity distance of the source $D_L$ (see e.g.  \citep{Holz2005} for more details).
For GW170817, it was measured $M_c=1.188^{+0.004}_{-0.002}$ M$_{\odot}$ and $i<55-56$ deg from which $D_L=40^{+8}_{-14}$ Mpc \citep{Abbott2017gw170817}. 

The possibility of measuring $D_L$ is a general property of all CBCs that, for this reason, are defined as 'standard sirens' in analogy with the 'standard candles' \citep{Schutz1986}. 
The detection of the electromagnetic counterpart of GW170817 and the subsequent sky position refinements, revealed the  host galaxy of this source, NGC4993, at z=0.009783 \citep{Hjorth2017}. The knowledge of both $D_L$ and $z$ is fundamental to constrain cosmological parameters. For nearby sources as GW170817, $D_L$ and $z$ yields the Hubble constant $H_0=cz/D_L$. For GW170817, it was found $H_0=70.0^{+12.0}_{-8.0}$ km s$^{-1}$ Mpc$^{-1}$. The $H_0$ measurement accuracy with just one source is still very low, though the obtained value is consistent with past measurements.
More importantly, this first result demonstrates that future multi-messenger observations of 'standard sirens' hold the promise to obtain an accurate, completely independent new measure of $H_0$, possibly solving the current tension among the measurements obtained using different probes \citep{Abbott2017H0}.


Contrary to the GW signal that was perfectly in agreement with predictions of a NS-NS merger according to General Relativity, the short GRB 170817A was peculiar in that its intrinsic luminosity $\sim10^{47}$ erg s$^{-1}$ is several orders of magnitude fainter than the average for short GRBs \citep{Abbott2017grbgw}. Also, the afterglow emission showed an anomalous slow rising behaviour, at odds with the typical decaying flux. The first afterglow detection in X-rays for GRB 170817A was obtained with Chandra only 9 days after the merger \citep{Troja2017}, i.e. at an epoch when typical X-ray afterglows are faded away. Extensive multi-band data analysis converged in finding both the prompt and the afterglow properties to be consistent with the first GRB viewed from outside the core of the jet (i.e. the cone with very high Lorentz factor), with viewing angle $\sim$15-30 deg from the direction of propagation of the highly relativistic jet core, a proxy of the inclination angle $i$ \citep{Ghirlanda2019,Mooley2018}. 
By including the information on the viewing angle in the luminosity distance computation, refined $H_0$ measurements could be obtained \citep{Guidorzi2017}, showing the powerful interplay between the GW and EM data. 
The EM follow-up of GW170817 also allowed us to detect and deeply characterize the 'kilonova' emission with unprecedented high resolution (e.g. \citep{Pian2017,Smartt2017,Coulter2017}). The 'kilonova' emission is a thermal component in the optical/NIR, with luminosity of the order of $1000$ times the nova ones (from which 'kilonova') that is predicted to develop after a few hours, and up to days after the merger of a NS-NS system (e.g. \citep{Li1998KN,Metzger2010}). Indeed, the neutron-rich  dense ejecta released after the NS-NS merger is predicted to have the required physical properties to trigger rapid neutron capture nucleosynthesis (r-process) forming unstable radioactive isotopes. The radioactive decay of such newly formed heavy elements heat the ejected material that, during its expansion, decreases its large opacity,  ultimately becoming transparent, and radiating thermal emission in the optical/NIR bands. However, it has been challenging to perform robust kilonova identification in optical /NIR surveys, even in combination with the detection of short GRBs. The temporal and spatial information on the NS-NS merger GW170817 and its proximity to the Earth (40 Mpc), enabled us to unambiguously find the kilonova component, identified with the optical transient source AT2017gfo, and to obtain unprecedented high-quality spectra at different epochs \citep{Pian2017,Smartt2017}. 
 Kilonova observations not only witness cosmic nucleosynthesis of heavy elements, but can also probe the physical conditions of the ejecta during and after the merger phase as e.g. the ejected mass and expanding velocity.

Since the end of the last observing run of aLIGO and AdV (March 2020), no other short GRBs was found to be associated with any detected CBCs, despite another GW event that was consistent with a NS-NS merger (GW190425) beside GW170817, and two events with NS-BH mergers \citep{Abbott2021nsbh}. A deep search for sub-threshold CBC GW events in coincidence with observed short GRBs was conducted for all the observation periods of the network aLIGO/AdV.  
This search was performed on GW archival data by assuming that the GW event sky position was consistent with the short GRB sky localization and by assuming that the binary merger happened within a temporal window of [$T_{GRB}$-5, $T_{GRB}$+1] s where $T_{GRB}$ is the GRB trigger epoch. The non-detection of any GW event in coincidence with a short GRB resulted in an estimate for the joint GRB+GW detection rate of $\sim$1/yr for the next scientific run (O4, currently planned to start by the end of 2022) \citep{Abbott2017GRBsearch}\citep{Abbott2019GRBsearch} \citep{Abbott2021GRBsearch}.

With the second generation (2G) GW detector network  current estimates are  $320^{+490}_{-240}$ Gpc$^{-3}$ yr$^{-1}$ for NS-NS systems \citep{Abbott2021ApJ913L7A} and $45^{+75}_{-33}$ Gpc$^{-3}$ yr$^{-1}$ for NS-BHs \citep{Abbott2021nsbh}. During the next decade (2030s), the advent of the next generation of GW interferometers, such as the Einstein Telescope \citep{Punturo2010-ET} and the Cosmic Explorer \citep{Reitze2019}, will extend the detection horizon up to and above the peak of star formation ($z\sim2$), with an expected detection rate of NS-NS mergers of the order of O($10^5$) per year (e.g \citep{Maggiore2020,Evans2021_CE}). With such a high detection rate, we expect the majority of short GRBs will be accompanied with GWs. A large sample of CBCs with an EM counterpart will allow us to make a very significant progress in our understanding of the physics of these objects, as well as to make relevant advances in cosmology. Table \ref{tab:table1} summarizes a few examples of fundamental questions that the combination of GW and short GRB observations will allow us to investigate. 

\begin{table}
\centering
\footnotesize
\begin{tabularx}{330pt}{|p{3cm}|p{8.3cm}|}
\hline
\textbf{Questions} & \textbf{Method} \\
\hline
How frequent is relativistic jet formation in NS-NS and NS-BH mergers? & The association of a short GRB with NS-NS/NS-BH mergers unambiguously marks the formation of a relativistic jet. Along with detections, confident non-detections of short GRBs in case of face-on mergers (orbital inclination angle can be extracted via the GW signal analysis) will provide the fraction of NS-NS and/or NS-BH with inefficient or no relativistic jet formation. \\
\hline
 What is the jet launching
mechanism in NS-NS/NS-BH mergers? & The time delay between the GW merger epoch and the GRB peak flux
are powerful diagnostic indicators for the jet launching mechanism (e.g. \citep{Abbott2017grbgw}). A significant number of short GRBs observed in synergy with GW detectors will allow us to uniquely characterize this important parameter and highlight possible differences between NS-NS and NS-BH systems. \\ 
\hline
 What is the nature of the NS-NS central remnant and the origin of the still unexplained short GRB X-ray extra-features (e.g. 'Extended Emission,'Plateaus')? & For short GRBs the monitoring of the subsequent X-ray emission in presence (or absence) of coincident continuous GW detection, will shed light on the nature of the merger remnant (i.e. BH, long-lived NS, other more exotic compact objects). This information will also unveil the origin and statistical properties of puzzling X-ray features like the 'Extended Emission' and the X-ray 'plateaus'.\\ 
\hline
Do jets have a universal structure and are there any systematic differences between NS-NS and NS-BH mergers? & The afterglow properties of short GRBs viewed from outside the core of the jet strongly depend on the jet structure. By knowing the progenitor system from GW data analysis, the detection several misaligned short GRBs and their long monitoring with powerful X-ray telescopes, will allow us to investigate on possible jet structure universality among NS-NS and NS-BH systems. \\
\hline
What is the role of NS-NS and NS-BH merger systems in the chemical enrichment of the Universe? & Kilonova emission encodes crucial information on the nucleosynthesis of the most heavy (r-process) elements. Kilonovae can be detected as transient optical/NIR sources by current and future surveys only up to a few $\times$100 Mpc and the search process can be severely challenged above $\sim100$ Mpc by the large number of optical/NIR transient sources expected in the large GW sky localization areas and the severe incompleteness of galaxy catalogs. 
X-ray and gamma-ray facilities can pinpoint the high-energy EM counterpart and ease the identification of a large number of kilonovae and their subsequent characterization through deep observations with sensitive optical/NIR telescopes.\\
\hline
At which rate the Universe is expanding? & A large sample of NS-NS/NS-BH with their EM counterparts will allow to provide an new, independent measure of the Hubble constant $H_0$ with the requested accuracy to solve the current tension between the values obtained using different probes.\\
\hline
\end{tabularx}
\caption{Some of the fundamental issues on the nature of NS-NS/NS-BH sources and short GRB that large sample of joint GW and EM detections with current and future high-energy missions will allow us to solve in synergy with the current and next generation GW interferometers.\label{tab:table1}}
\end{table}

\subsubsection{Joint Neutrino and EM observations of GRBs}

In the most accepted paradigm, GRBs are generated via synchrotron radiation from non-thermal electrons that are shock-accelerated inside a jet. Besides electrons, ions can also be accelerated in the jet, and high-energy neutrinos can be produced by p$\gamma$ interactions. Neutrinos in the energy range of 0.1-1 PeV are expected to emerge from the decay of charged pions, as well as from muon decay
(e.g. see  \citep{Murase2019} for a review). 

Despite the 0.1-1 PeV band having the best sensitivity for what is now the most powerful neutrino telescope (i.e. IceCube), so far no neutrino events coincident with bright GRBs have been found. This is in spite of several coordinated searches within the temporal window of GRBs. This shows that neutrino production efficiency in bright GRBs is much lower than theoretically expected and severely limits the role of GRBs for the astrophysical neutrino fluxes observed with IceCube \citep{Aarsten2013}, as well as for the ultra-high energy cosmic ray (UHCRs).  

However, neutrino searches have so far been focused on the brightest GRBs, since they are thought to be associated with the formation of the most powerful relativistic jets and thus with highly efficient hadron acceleration mechanisms. 
Low-luminosity GRBs (LL-GRBs, e.g. \citep{Virgili2009}) form a subclass of long GRBs that shows both peculiar low luminosity values around $L_{iso}\sim10^{49}$ erg s$^{-1}$, and that it is thought to be up to three orders of magnitude more numerous than standard long GRBs. LL-GRBs have been interpreted as produced by mildly relativistic jets with Lorentz factors $\Gamma$ much lower than the standard, bright long GRBs. Low $\Gamma$s can be due to a large baryonic content of the jet or to jets that loose large part of their energy content while carving the dense stellar envelope (nearly chocked jets, see e.g. \citep{Fasano2021,Zhang2018grb_uhcr}). 
In such conditions, while gamma-rays suffer from strong absorption, which render our classification of the source as low-luminosity, neutrino emission do not. Given the predicted larger rate of LL-GRBs with respect to standard bright long GRBs and the expected larger number of p$\gamma$ interactions, this class of sources have been suggested to be more promising neutrino sources to be addressed in  future multi-messenger observations (e.g. \citep{Kimura2022}). 

In the case of short GRBs, proton-proton collisions in the accretion disk of NS-NS mergers are also expected to take place and contribute to the neutrino emission. The nearby short GRB 170817A was observed at large off-axis angle and in a non-optimal position with respect to the operating neutrino observatories. For these reasons, the lack of any coincident neutrino event (e.g. \citep{Albert2017}) did not lead to any robust conclusion. 
Some works have suggested that high-energy neutrinos can be efficiently produced during the “Extended Emission" (EE) of short GRBs, i.e. the $\sim$100s lasting soft gamma-ray emission following a fraction of short GRBs.  Among the possible interpretations of the mysterious origin of the EE, there is a slower and more isotropic ejecta with lower Lorentz factor. Being analogous to LL-GRBs, this component may represent an efficient neutrino source \citep{Kimura2017}.

\subsection{Blazars}

Blazars are classified as the most powerful sources of persistent non-thermal radiation in the Universe (e.g. \citep{Urry1995}). They are a particular type of Active Galactic Nuclei (AGN), with highly variable emission over the entire electromagnetic spectrum. Radiation from blazars originates from a relativistic jet that moves away from the central supermassive black hole, and is oriented in the direction of the Earth. This is the very special condition that is responsible for the distinctive features of these sources such as superluminal motion and fast variability (e.g. \citep{Giommi2021}). 

A few thousand blazars are known so far and typically classified in two flavours: Flat-Spectrum Radio Quasars (FSRQs) and BL Lacertae objects (or BL Lacs). FSRQs show broad emission lines in their optical spectrum while BL Lacs are featureless or with very weak emission lines (e.g. \citep{Falomo2014}.


Blazars' $\nu$f($\nu$) spectral energy distribution shows two broad humps, the first in the range from IR to X-rays, the second in the $\gamma$-rays. 
In the standard ‘leptonic' scenario, the low- and high-energy spectral peaks are explained by synchrotron emission and inverse-Compton radiation from non-thermal electrons, respectively. For blazars with low luminosities, seed photons for the inverse-Compton scattering are mainly provided by the electron synchrotron process. On the other hand, in the case of more luminous blazars, such as the flat-spectrum radio quasars (FSRQs), seed photons have more likely an external origin, e.g. from the accretion disk or the broad-line regions. In the 'lepto-hadronic' scenario, gamma-rays originate from hadronic particles induced electromagnetic cascade emission or ion synchrotron radiation. In the former case, comparable neutrino and gamma-ray fluxes are expected, while in the ion synchrotron radiation scenario strong magnetic fields are required and neutrino production efficiency may be low.

\subsubsection{Joint Neutrino and EM observations of blazars}

Blazars represent the first astrophysical sources at cosmological distances from which neutrinos have been detected, with so far only one case. Indeed, in September 2017, IceCube detected a muonic neutrino of $\sim290$ TeV (IC170922) and sent an alert to the astronomical community providing the time of the detection and the sky localization \citep{Kopper2017}. The Cherenkov Telescope MAGIC revealed a gamma-ray signal temporally consistent with IC170922, and identified the source with the flaring blazar TX506+056 at z=0.3 \citep{Aartsen2018}. TX506+056 is the first electromagnetic counterpart of a neutrino event found at cosmological distances, i.e. beyond the Local Group, and it is an important indication of the possible role of blazars in explaining the origin of cosmic neutrino flux observed with IceCube \citep{Aarsten2013}.

Interestingly, the EM/$\nu$ properties of TX506+056 challenge the theoretical predictions from these objects being the lepto-hadronic scenario not able to explain both the gamma-ray and the neutrino event observations. In fact, the observed gamma-ray emission is incompatible with hadronic particles induced electromagnetic cascade emission that, on the contrary, would naturally explain the IceCube observations. On the other hand, the blazar spectral energy distribution can be modelled by invoking proton synchrotron emission, but in this case the IceCube observations cannot be simultaneously explained. A viable solution to these inconsistencies is the hypothesis that the two messengers originated from different regions, matching the so called 'multi-zone' model \citep{Keivani2018,Murase2019}.

\subsection{Other multi-messenger source candidates}

In this section we summarize other classes of high-energy  sources that are believed to be candidates for multi-messenger observations. Despite the large uncertainties in the theoretical predictions of their detectability, these sources will most likely be observed with the next generation of neutrino and GW detectors during the 2030s.  

\subsubsection{Core-collapse SNe: Long GRBs and Shock Breakouts}

SN shocks are the preferred locations where the CRs are believed to be accelerated and neutrino events generated. To this regard, the next generation of neutrino observatories will be fully exploited in the detection of MeV neutrinos emerging from the very first stage of the SN explosion. Pulses of low energy neutrinos ($<10$ MeV) are expected to be released during core collapse (cc)-SNe with an energy release of up to $10^{53}$ erg. So far, MeV neutrinos have been detected from one cc-SN, namely SN1987A, in the Large Magellanic Cloud at a 50 kpc distance (e.g. \citep{Koshiba1992}). 
cc-SNe are also candidate for high-frequency GW sources if a sufficiently high level of asymmetry is present during the explosion. However, contrary to the CBC case, their expected GW emission is highly uncertain as it strongly depends on the rather unknown SN explosion mechanism. While this makes it difficult to predict the GW signal and its detectability, it represents a unique opportunity to probe the cc-SN inner dynamics, inaccessible to electromagnetic observations. Both GWs and neutrinos can provide important information on the degree of asymmetry in the matter distribution, as well as the rotation rate and the strength of the magnetic fields (e.g. \citep{Powell2022}). 

Shock breakouts (SBOs) represent the soft X-ray counterpart of ccSNe, expected to follow closely ($\sim$10-1000 s) the core-collapse time. In particular, SBOs from Wolf–Rayet stars, as well as red and blue supergiants, are expected to appear as bright X-ray bursts lasting between 10–1000 s and with luminosities in the range $10^{43}–10^{46}$ erg s$^{-1}$. Such stars are the likely progenitors of Type Ibc and most Type II SNe. 
However, so far the only robust evidence of a shock-breakout form a ccSN is SBO080109 (25 Mpc) serendipitously detected with Swift/XRT \citep{Soderberg2008}. 
SBOs are temporally closer to the possible GW/neutrino signal produced during the collapse than the optical cc-SNe counterparts. As a consequence, their detection potentially marks the start time of the GW/neutrino event with more accuracy with respect to the SN signal, ultimately allowing for more effective GW/neutrino event searches in archival databases
(e.g. \citep{Andreoni2016}).

More promising GW signals from cc-SNe may originate from the compact object formed in the core collapse, soon after birth. Different models were also proposed also for this scenario, which rely on the formation of a millisecond spinning NS and the release of its spin energy reservoir $>10^{-2}$ M$_{\odot}$ c$^2$ i.e. comparable to NS-NS mergers (e.g. \citep{Corsi2009,Dallosso2018}. The expected rates in this case depend on the fraction of NS that are formed with the ms-spin required to produce copious continuous GW emission. During the 2G network era, one may expect the detection of a cc-SN signal from within 100-200 kpc, or up to $\sim$2-3 Mpc if a ms-spinning, likely highly-magnetized NS, was formed in the core-collapse. Next generation GW detectors, with their $\sim10$ times increased sensitivity, will lead to a corresponding extension of the expected horizon ($\sim20-30$ Mpc). The joint detection of the GW signal from a ms-spinning NS, likely a magnetar (a NS with the highest magnetic fields, e.g. $10^{14-15}$ Gauss), and of the EM signatures of its birth would represent ‘‘per se'' a breakthrough discovery for NS physics. 

Long GRB phenomena mark the core-collapse of a particular type of massive stars characterized by a large angular momentum, and for which no hydrogen, and possibly helium, external envelopes are retained at the time of the explosion. The latter property is inferred from the absence of hydrogen and in some cases also of helium spectral lines in the optical spectra of the associated supernovae (SNIb,c). 
Any long GRB, if close enough, could be associated with a detectable GW emission and thus offers a very interesting potential synergy between gamma-ray and GW detectors. The first joined GW/GRB/SN observations, possibly also combined with neutrino detections, will prove crucial to unravel the nature of these sources and their explosion mechanism.
Observed long GRB rate densities are of the order of O(1) Gpc$^{-3}$ yr$^{-1}$ (e.g. \citep{Lan2019}). Thus, the simultaneous GW+EM detection is not expected before the next (third) generation GW interferometers. The predicted large number of LL-GRBs in the nearby Universe, expected to be up to O(10$^3$) times more numerous than long GRBs, are the best candidates for EM+GW detection, not only because of their larger population, but also because of their much smaller distances with respect to long GRBs. Interestingly, LL-GRBs are also preferred targets for coincident neutrino searches with respect to bright long GRBs (see previous sections), thus making this class of sources very promising multi-messenger targets for the near future. 

\subsubsection{Bursting magnetars and Soft Gamma Repeaters}

The most widely accepted explanations to interpret the magnetar bursting activity, and in particular the rare giant flares observed in X-rays from soft gamma repeaters (SGRs; see e.g. \citep{Mereghetti2015}), invoke the fractures of the solid crust on the surface, or dramatic magnetic field readjustments in highly magnetised NSs. These events will inevitably excite non-radial oscillation modes that may produce detectable high-energy GWs (see e.g. \citep{Corsi2011,Ciolfi2011}). The most recent estimates for the energy reservoir available in a giant flare are between $10^{45}$ erg (about the same as the total electromagnetic emission) and $10^{47}$ erg. The efficiency of conversion of this energy into GWs was estimated in numerical relativity simulations, and it was found to be likely too small to be within the sensitivity range of present GW detectors \citep{Ciolfi2012}. However, at the typical dominant (i.e. f-mode) oscillation frequencies in NSs ($\sim$kHz), next generation ground-based GW detectors will be sensitive to much lower GW energies \citep{Punturo2010-ET}. Therefore, a relatively close giant flare event might lead to a detectable GW emission.

\begin{figure}
    \centering
    \includegraphics[scale=0.5]{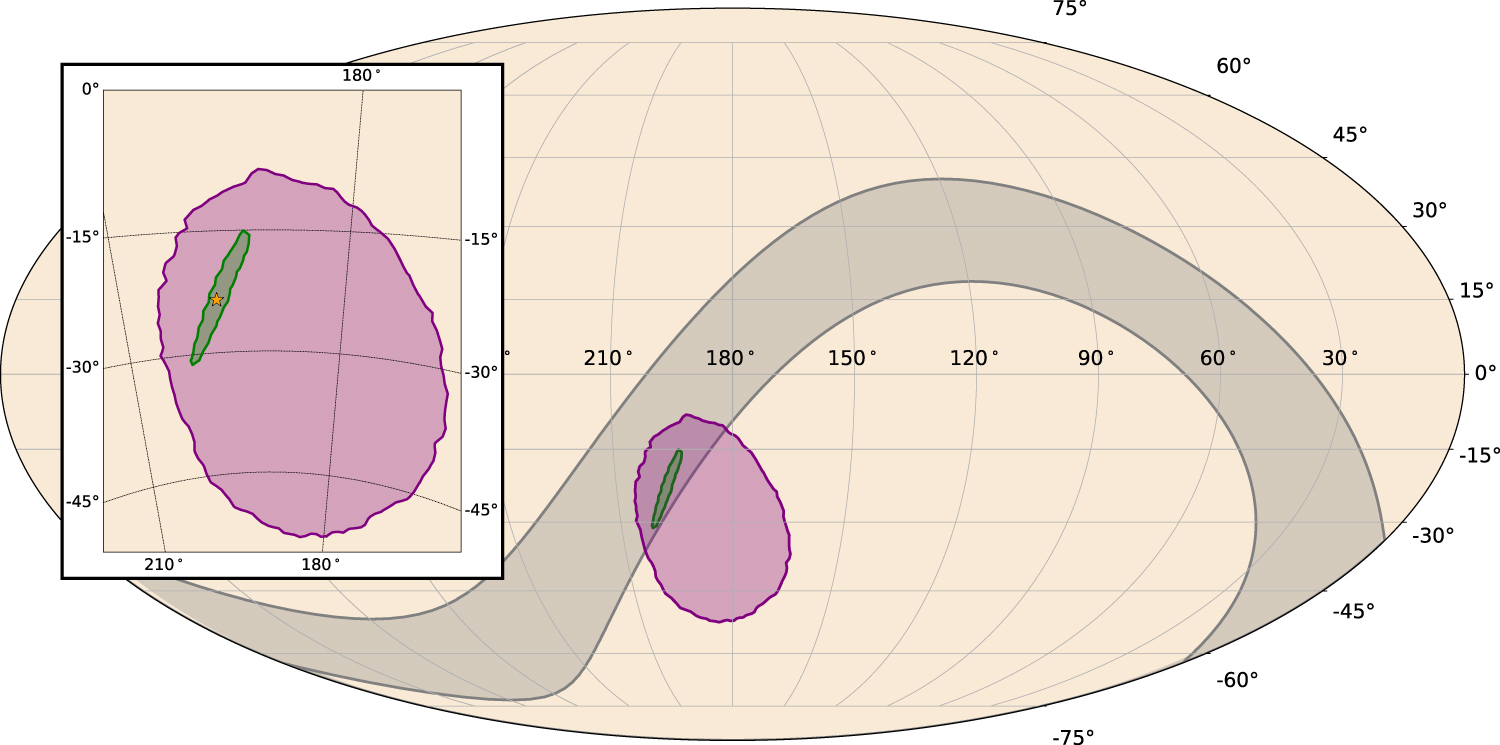}
    \caption{Sky localization of the NS-NS merger GW170817 detected with aLIGO and AdV, and of the associated GRB 170817A independently detected with the high-energy  mission Fermi and INTEGRAL \citep{Abbott2017grbgw}. The simultaneous sky coverage of Fermi and INTEGRAL allowed us to detect the associated electromagnetic counterpart and to confirm the spatial and temporal association with the GW event. Figure from Ref. \citep{Abbott2017grbgw} under the terms of the Creative Commons Attribution 3.0 License.}
    \label{fig:skyloc}
\end{figure}

\section{\textit{Multi-messenger observations}}

In the previous sections, we have described the enormous utility of joint GW/neutrino/EM observations. At the same time, such combined observations have been proven to be extremely challenging in the past years due to several factors, among which the insufficient sensitivity of neutrino and GW detectors, the poor sky localization of GW events  and of many neutrino events, and the transient nature of a large part of multi-messenger sources that can be detected with the available facilities. 

In this section, after a description of the main GW and neutrino facilities which will be used by multi-messenger observational campaigns in the next years, the crucial role of high-energy surveys for detecting the EM counterparts of GW and/or neutrino events is highlighted, together with the main X-ray and gamma-ray facilities that will contribute to multi-messenger follow-up campaigns during the 2020s and the 2030s.

\subsection{High-frequency gravitational wave detectors}

The long standing problem in detecting GWs has been due to instrument  sensitivities, which for several decades was insufficient to catch the expected tiny signals from space-time perturbation due to the crossing of GWs.  Indeed, GWs have been searched for nearly one century with no success. Nevertheless, this search stimulated breakthrough technological progresses, particularly in the field of laser interferometry, that culminated in the 2015 detection of GW \citep{LVC-BBH1}. 

GWs distort space-time by changing distances between freely falling objects 
in the plane perpendicular to the direction of propagation. The crossing of GWs make distances shorter in one direction, and longer in the perpendicular direction, and for this reason Michelson interferometers, which measure the differences in length between perpendicular arms, have been used and adapted to measure the effect of a crossing GW. 
However, such measurements are extremely challenging. To give an example, a NS-NS system with NS separated by 200 km and at a distance of 20 Mpc from Earth, would emit GWs with strength $h\sim10^{-22}$ at $\sim300$ Hz (e.g. \citep{Gonzalez2007}). 
For an interferometer with km-length arms, such strength $h$  corresponds to a distance change $\Delta L=Lh$ of the order of O($10^{-19}$) m.  

Decades of experimental efforts on laser interferometry to detect GWs were awarded in September 2015 with the first detection \citep{LVC-BBH1} and then in 2017 with the Nobel Prize for Physics. 
aLIGO \citep{aLIGO} represents the second generation of GW interferometers and it is composed by a network of two detectors located in Hanford (WA) and Livingston (LA) in the USA. aLIGO has 4-km length arms and is about one order of magnitude more sensitive to GW in the frequency band $\sim20-2000$ Hz than the first generation. 
For more details on aLIGO detectors we refer the reader to \citep{aLIGO} and references therein. 

Advanced Virgo (\citep{AdVirgo}) is another GW interferometer of second generation located in Italy that joined the aLIGO network in August 2017. AdV has 3 km-length arms. The name ‘‘Virgo'' comes from the galaxy cluster at $\sim20$ Mpc from which GW signals were expected to be detected with high probability given the cluster proximity and the large number of galaxies it contains. AdV has an improved sensitivity of one order of magnitude with respect to the initial Virgo. For more details on Virgo detector we refer the reader to \citep{AdVirgo} and references therein. 

In February 2019, a fourth interferometer joined the aLIGO/AdV network: KAGRA, from the Japanese word KAGURA, which is the abbreviation of KAmioka GRavitational-wave Antenna (but also the word for a traditional sacred music and dance for the gods). KAGRA has 3-km length arms but, in contrast to AdV and aLIGO, it is located underground to be isolated from the seismic motion. Another difference is in the core optics being cooled down to 20 K to reduce thermal fluctuations. We address the reader to \citep{Somiya2012,Aso2013} for more details on KAGRA characteristics.

In $\sim$2025, both aLIGO and AdV will reach a sensitivity larger than the nominal one with the upgraded configurations ‘‘A+'' and ‘‘Virgo+'' \citep{Abbott2020LRR} during the fifth observation run (O5, the start of which is currently scheduled for mid 2025). By that time, the network of GW interferometers with comparable sensitivity, will be able to detect neutron star binary mergers up to $z\sim0.1$, with an expected rate of few tens of detections per year. By the end of the twenties, a new gravitational wave detector located in India (LIGO-India) is planned to join the network \citep{GWIC2020}. 
At the same time, major upgrades are under consideration for aLIGO/A+ in the ‘‘Voyager'' configuration \citep{Reitze2019} that will test some of the key technologies needed for the third generation (3G) interferometers, such as the Cosmic Explorer (CE, \citep{Reitze2019,Abbott2017nextgen}) and Einstein Telescope (ET, \citep{Punturo2010-ET}, \citep{Hild2011}\citep{Maggiore2020}) expected to be operational by mid-thirties. The ”Voyager” configuration will possibly place the GW detection rate mid-way between the 2G and 3G detectors by the end of the twenties or early 2030s. The 3G detectors will have about one order of magnitude better sensitivity than current ones, with an expected rate of compact binary coalescences (CBCs) per year of the order of O($10^5$), up to distances that will cover the peak of star formation epoch and far beyond \citep{Maggiore2020,Evans2021_CE}.

A major issue for multi-messenger observations concerns the GW source sky localisation accuracy. The latter is based mainly on the triangulation method that exploits the temporal delay in the GW signal detection among different interferometers, and so it is strongly dependent on the number of detectors forming the network. A Bayesian approach is typically followed, where the posterior probability distribution for the position in the sky of the GW source is constructed using the signal arrival time at each detector, as well as the consistency between the phase and amplitude of the GW signal among different detectors (see \citep{Abbott2017nextgen} and references therein). 

By the end of the twenties, the completed GW detector network will have improved sky localisation capabilities, with average uncertainty regions of the order of few tens of square degrees for CBCs \citep{Abbott2020LRR}. 
During the thirties, the GW source localisation will still rely mostly on the triangulation method and therefore will depend on the presence of a network of 3G interferometers \citep{Yufeng2022,Borhanian2022,Ronchini2022}. 
In the local Universe, however, the high signal to noise ratio with which CBC events will be detected will make it possible to have some sky localization capabilities even if they work separately. The slight extension of their sensitivity to lower frequencies below 10 Hz will permit to detect the inspiral phase of CBCs well in advance the merger epoch. 
For example, ET will detect NS-NS systems up to a few hours before the mergers. This will allow us to exploit the effects of Earth rotation on the signal in the GW source sky localization computation. ET alone will be able to detect a few hundreds of NS-NS mergers per year with sky-localization less than $\sim$100 square degrees \citep{Ronchini2022} up to redshift $z\sim0.8$ while a network including ET and CE will detect thousands of events with sky-localization less than 10 square degrees. 
At larger distances ($z>0.8$), the sky localization accuracy decreases to several tens to hundreds of square degrees for the large majority of the events \citep{Ronchini2022}.


Early warnings from GW detectors are pre-merger alerts sent when the inspiral GW signal is entering in the sensitivity band of the detector \citep{Nitz2021}. Early warnings are extremely important for the astronomical community, since they allow ground and space-based telescopes to slew onto the identified sky region before the advent of the expected electromagnetic transient emission. Early warning alerts of the order of few minutes for the most nearby sources (i.e. NS-NS mergers at few tens of Mpc) can be obtained with the 2G interferometers \citep{Sachdev2020} and can be as early as hours with 3G detectors (e.g. \citep{Chan2018,Nitz2021}).

\subsection{Neutrino detectors}

The low cross sections of neutrinos make them powerful messengers for the most dense and/or distant regions of the Universe, but at the same time render their detection extremely challenging. 
Large volumes of the order of $\sim1$ km$^3$ are required, and natural media such as water or ice allowed us to solve the technological challenges of such volumetric facilities. By interacting with water or ice, a neutrino produces a muon or another charged lepton, depending on the neutrino flavor. The newly formed lepton propagates at a velocity higher than light speed in the medium, generating Cherenkov light. The latter is detected by photomultipliers distributed all over the detector volume, ultimately providing the neutrino energy and direction of origin in the sky. Ice and water also act as an atmospheric muon screen that represent the first background source. Deep regions of ice or water neutrino detectors of about 1-2 km allows us to reduce the atmospheric muon flux by a factor of $10^6$. However, given the low flux of cosmic neutrinos, they are typically identified when coming from the other side of the Earth rather than directly overhead in the sky, i.e. from underground, since only neutrinos can cross the Earth. Therefore, to guarantee an all sky monitoring, at least two comparable sensitivity neutrino detectors (one per hemisphere) are required. 

IceCube at the South Pole is operative since 2008 in its complete configuration, and it is composed by 86 columns of 60 fotomultipliers each, for a total of 5160 multipliers and a volume of 1 km$^3$. 
The Km3NeT \citep{Adrian2016} is under construction at 3800 mt below the Mediterranean Sea and it will have a volume of the order of 1 km$^3$. Km3NeT will provide IceCube comparable sensitivity in the Northern Hemisphere on the second half of the twenties and it will work in synergy with another detector in the Northern Hemisphere, the Gigaton Volume Baikal lake neutrino detector.
IceCube-gen2 is the next generation neutrino telescope and it is conceived as an extension of the IceCube detector volume. IceCube-gen2 construction phases will gradually increase the current IceCube sensitivity up to one order of magnitude (IceCube-Gen2 Collaboration, \citep{Aartsen2014}), boosting the current astrophysical neutrino detection rate to values that will significantly increase the chances to find the electromagnetic counterparts during the 2030s. 

The neutrino sky localization accuracy depends on neutrino flavour, where long tracks topology for $\nu_{\mu}$ provide angular resolution down to 0.1-0.2 deg and a 2$\pi$ sr sky coverage, while cascade topology for $\nu_e$ and most $\nu_{\tau}$ angular resolution are typically 3-5 deg for the whole $4\pi$ sky. Thus, in order to unambiguously identify the neutrino source, cross-correlations with sky surveys in the electromagnetic spectrum are mandatory. 

\begin{figure}
    \centering
    \includegraphics[scale=0.5]{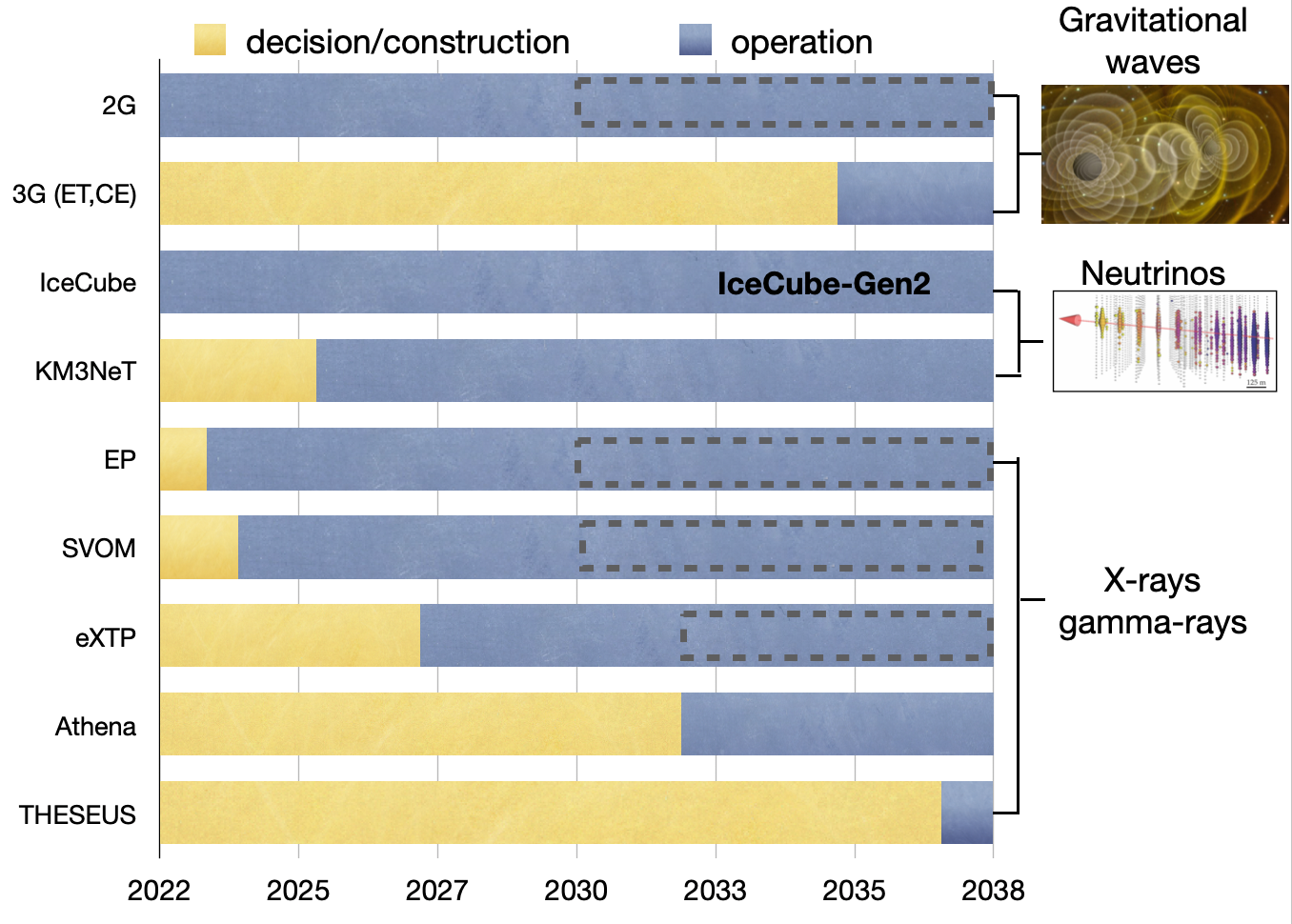}
    \caption{Tentative timeline for current and future ground-based GW interferometers, neutrino detectors and high-energy facilities that are expected to operate in the next multi-messenger observational campaigns. Yellow regions indicate the period before the operational one, the latter being coloured in blue. Dashed lines mark possible extension.}
    \label{fig:timeline}
\end{figure}

\subsection{X-ray and gamma-ray facilities}


The two breakthrough results obtained with the first detection of the electromagnetic counterparts of a GW source (GW170817/GRB 170817A) and of a neutrino event (IC170922/TX506+056) described in the previous sections, clearly demonstrated the crucial role of high-energy observations in the multi-messenger observational campaigns.
There are a number of advantages in performing multi-messenger follow-up at high energies with respect to the lower ones, for instance:
\begin{itemize}
\item At distances larger than $\sim$100 Mpc, even a relatively accurate sky localization from GW or neutrino detectors of the order of 0.1-1 square degrees can challenge the identification of the electromagnetic counterparts at optical wavelengths. The reason is the large number of transient sources expected in such sky areas at large distances and the severe incompleteness of galaxy catalogs above 100 Mpc. On the contrary, the high-energy transient sky is much less populated and gamma-ray or X-ray counterparts detected with all sky monitors might represent the most suitable way to pinpoint the GW source.
\item In several cases, GW/neutrino sources will be localized through their gamma-ray counterparts with an accuracy that will still not be enough to unambiguously identify the electromagnetic counterpart using optical observations. X-ray narrow field instruments, such as XMM, Chandra, eXTP, Athena, will provide the required refinement on the sky localization down to the arcmin/arcsecond level. 
\item X-ray and gamma-ray transients are typically better tracers of the epoch of GW or neutrino emission with respect to longer wavelength counterparts, being the high-energy emission typically closer in time to the expected epoch of GW or neutrino emission. Accurate estimate of the GW or neutrino event epoch, together with the knowledge of the sky localization from X-ray/gamma-ray observations, significantly increases the chances to detect GW/neutrino sub-threshold signals in the so called triggered searches. 
\end{itemize}

Hereafter, we provide a brief description of the main X-ray and gamma-ray missions that are expected to be launched in the next years, and that will provide an important contribution to multi-messenger follow-up campaigns in synergies with several current and future facilities operating at longer wavelengths such as the Rubin Observatory, the Very Large Telescope, the Extremely Large Telescope, the Square Kilometer Array, etc. 

\subsubsection{Einstein Probe}
The Einstein Probe (EP) is a small mission dedicated to monitor the sky in the soft X-ray band led by the Chinese Academy of Sciences with international collaboration. It will carry out systematic survey and characterisation of high-energy transients with two detectors: a wide-field X-ray Telescope (WXT, 0.5-4 keV) with $60\times60$ deg$^2$ Field of View (FoV) and spatial resolution of $\sim5$ arcmin) and a narrow-field Follow-up X-ray Telescope (FXT, 0.3-10 keV, $1$ deg$^2$ FoV), as well as a fast alert downlink system. To achieve both a wide FoV and X-ray focusing, the novel Micro-Pore Optics (MPO) in the lobster-eye configuration is adopted for WXT \citep{Yuan2015_EP}. 

EP scientific goals focus on discovering X-ray transients as tidal disruption events, supernova shock breakouts, soft-faint GRBs as those at high-redshifts or the low-luminosity GRBs, and new types of transients, among which electromagnetic counterparts of gravitational wave and neutrino events are expected to be observed with the currently available 2G GW interferometers and neutrino telescopes. 
The mission is scheduled for launch by the end of 2022 \citep{Liu2021_EP}. 

\subsubsection{SVOM}
The Space-based multi-band astronomical Variable Objects Monitor (SVOM)  focuses on the detection of GRBs and is the most important GRB-dedicated mission next to the currently operating Niel Gehrels Swift mission \citep{Gehrels2004-Swift}, with multi-band observation and fast slew capabilities. 
In addition to Swift, SVOM will have dedicated ground-based optical/NIR telescopes for follow-up observations. SVOM is an international cooperation project led by the Chinese National Space Agency (CNSA) and the Centre National d’Etudes Spatiales of France (CNES), and is composed as follows:
\begin{itemize} 
\item Two wide-field high-energy detection payloads: the hard X-ray camera (ECLAIR, 4-250 keV, $89\times89$ deg$^2$ FoV, $13'$ positional accuracy) and Gamma-ray monitor (GRM, 0.015-5 MeV, 60 deg FoV, $\sim5$ deg positional accuracy). 
\item Two narrow-field telescopes: the soft X-ray telescope (MXT, 0.3-10 keV, $64'\times64'$ FoV, $2'$ positional accuracy) and visible light telescope (VT, $0.4-1\mu$m, $26'\times26'$ FoV, 1 arcsec positional accuracy). 
\end{itemize} 

The SVOM/ECLAIR sky observations will be constantly monitored by ground-based telescopes, the ground wide-angle camera array (GWAC, $0.4-0.9\mu$m) which can cover a part of the ECLAIR’s viewing field in real time, and by two 1 mt class ground robotic follow-up observation telescopes (GFTs, $0.4-1.7\mu$m).  

Similarly to EP, SVOM will focus on low-luminosity and high redshift GRBs but it will provide multi-wavelength information and fast ground-based optical/NIR follow-up \citep{Wei2016-SVOM}. 
SVOM will provide accurate GRB position, allowing us to activate follow-up campaigns with other facilities. 
Additional targets include AGN and in particular blazars, that together with GRBs are the most promising  electromagnetic counterparts for the next years \citep{Atteia2019}. SVOM is currently planned to be launched by mid 2023 \citep{Yu2020SVOM}.

\subsubsection{eXTP}

The enhanced X-ray Timing and Polarimetry mission (eXTP) is a science mission designed to study the state of matter under extreme conditions of density, gravity and magnetism. The planned launch date of the mission is 2027.  

The payload is composed of:
\begin{itemize}
    \item The Spectroscopic Focusing Array (SFA), a set of 9 X-ray telescopes (0.5-10 keV, 12 arcmin FoV each). 
    \item the Large Area Detector (LAD), a deployable set of 640 Silicon Drift Detectors, achieving a total effective area of $\sim3.4$ m$^2$ between 6 and 10 keV. The operational energy range is 2-30 keV and it is a non-imaging instrument, with the FoV limited to $<1$ deg FWHM. 
    \item the Polarimetry Focusing Array (PFA), a set of 4 X-ray telescopes (2-10 keV, 12 arcmin FoV each), equipped with imaging gas pixel photoelectric polarimeters.
    \item the Wide Field Monitor (WFM), a set of 3 coded mask wide field units, equipped with position-sensitive Silicon Drift Detectors (2-50 keV, FoV of 3.7 sr).
\end{itemize}

Primary targets include isolated and binary neutron stars, strong magnetic field systems like magnetars, and stellar-mass and supermassive black holes, including the most promising multi-messenger sources so far that are short GRBs and blazars, as well as other potential candidates as long GRBs and the shock breakout from core collapse supernovae \citep{intZand2019eXTP_observatory_science}.  


\subsubsection{Athena}
Athena (Advanced Telescope for High ENergy Astrophysics) is the X-ray observatory large mission selected by the European Space Agency (ESA), within its Cosmic Vision 2015-2025 programme, to address the Hot and Energetic Universe scientific theme \citep{Nandra2013}, and it is provisionally due for launch in the early 2030s. 

Athena will have three key elements to its scientific payload: an X-ray telescope with a focal length of 12 m and two instruments: 
\begin{itemize}
    \item a Wide Field Imager (WFI, $40'\times40'$ FoV) \citep{Meidinger2020} for high count rate, moderate resolution spectroscopy over a large FoV 
    \item an X-ray Integral Field Unit (Athena/X-IFU, 5' effective diameter FoV, \citep{Barret2018} for high-spectral resolution imaging.
\end{itemize}

Athena will allow us to study the central engine and jet physics in compact binary mergers as well as the nucleosynthesis of heavy elements through the follow-up of the X-ray long-term emission. AGN, and in particular blazars, are also among the main targets of Athena. Given the transient nature of the most likely multi-messenger sources, the synergy of Athena with wide-field high-energy facilities will be fundamental \citep{Piro2021}. 

\subsubsection{THESEUS}
The Transient High-Energy Sky and Early Universe Surveyor (THESEUS) is a mission concept designed to exploit GRBs as cosmic probes of the Early Universe as well as multi-messenger sources. THESEUS is one of the three mission concepts that were selected by ESA in 2018 for a Phase A study as candidates for the medium-class  mission call M5. 
The THESEUS project is being further developed for responding to new opportunities for medium-class missions. The next launch schedule of these new opportunities is planned for the end of the 2030s when the scientific return of THESEUS for multi-messenger astrophysics will further improve by allowing for a great synergy with next generation GW and neutrino detectors in a consolidated operational phase (Fig.\ref{fig:timeline}).

The foreseen payload of THESEUS includes the following instrumentation:
\begin{itemize}
    \item The Soft X-ray Imager (SXI, 0.3 – 5 keV), a set of 2 lobster-eye telescopes units, covering a total FoV of $\sim$0.5 sr with source location accuracy $< 1-2$ arcmin;
    \item The InfraRed Telescope (IRT, 0.7 – 1.8 $\mu$m), a 0.7m class IR telescope with $15'\times15'$ FoV, for fast response, with both imaging and spectroscopy capabilities;
    \item X-Gamma rays Imaging Spectrometer (XGIS, 2 keV – 20 MeV), a set of 2 coded-mask cameras using monolithic X-gamma rays detectors based on bars of Silicon diodes coupled with CsI crystal scintillator, granting a $\sim2$sr FOV and a source location accuracy of $\sim10$ arcmin in the 2-150 keV, as well as a $>4$ sr FoV at energies $>$150 keV.
\end{itemize}

Soft GRBs, such as those at high redshift, as well as those expected to be seen off-axis as GRB 170817A and the low-luminosity GRBs will be best studied with THESEUS/XGIS and THESEUS/SXI, where the last two classes of GRBs are among the most promising electromagnetic counterparts of NS-NS mergers and core-collapse massive stars. Blazars are also among the main targets of THESEUS/XGIS. 

If selected, THESEUS will be launched by 2037. The end of the thirties will be characterized by a huge detection rate of compact binary coalescences with the third generation GW interferometers as ET and CE, as well as a high rate of neutrino events with IceCube-Gen2 and Km3Net. The role of THESEUS will therefore be crucial for the full exploitation of these powerful multi-messenger facilities \citep{Amati2021,Stratta2018,Ciolfi2021}.

\section{Cross-References}

\begin{itemize}
\item The Einstein Probe Mission
\item SVOM
\item eXTP
\item ATHENA
\end{itemize}




\bibliographystyle{abbrv}
\bibliography{All_refs.bib}

\end{document}